\documentclass[prx,onecolumn,nofootinbib,citeautoscript,eqsecnum,10pt,longbibliography,notitlepage,showkeys]{revtex4-2}

\usepackage{graphicx}% Include figure files
\usepackage{dcolumn}% Align table columns on decimal point
\usepackage{bm}% bold math
\usepackage{color} 

\usepackage[papersize={8.5in,11in}]{geometry}

\usepackage{color}
\definecolor{darkblue}{rgb}{0.,0.,0.4}
\definecolor{darkred}{rgb}{0.5,0.,0.}
\definecolor{BlueViolet}{RGB}{138,43,226}
\definecolor{SkyBlue}{RGB}{30,144,255}
\definecolor{DarkGreen}{RGB}{0,100,0}
\usepackage[pdftex,colorlinks=true,linkcolor=darkblue,citecolor=blue,urlcolor=darkred]{hyperref}

\usepackage{physics}
\usepackage{cancel}

\newcommand{\beq}{\begin{eqnarray}}
\newcommand{\eeq}{\end{eqnarray}}
\newcommand{\beqq}{\begin{eqnarray*}}
\newcommand{\eeqq}{\end{eqnarray*}}
\newcommand{\be}{\begin{equation}}
\newcommand{\ee}{\end{equation}}

\newcommand{\nn}{\nonumber \\}

\renewcommand{\vec}[1]{{\bf #1}}

\renewcommand{\epsilon}{\varepsilon}

\usepackage[utf8]{inputenc}
\usepackage{graphics}
\usepackage{epsfig}
\usepackage{times}
\usepackage[dvipsnames]{xcolor}
\usepackage{bbm,bbold}   
\usepackage{amsfonts}
\usepackage{amssymb}
\usepackage{amsthm}
\usepackage{bm} 
\usepackage{amsmath}
\usepackage{xr-hyper} 
\usepackage{hyperref}
\usepackage{color}
\usepackage{graphicx}
\usepackage{subfigure}
\usepackage{wasysym}
\usepackage{soul}
\usepackage{verbatim}

\begin{document}

\title{Magnetotropic Response in Ruthenium Chloride}

\author{Ipsita Mandal}
\email{ipsita.physics@gmail.com}
\affiliation{Institute of Nuclear Physics, Polish Academy of Sciences, 31-342 Krak\'{o}w, Poland}

\begin{abstract}
We consider the exchange couplings present in an effective Hamiltonian of $\alpha$-RuCl$_3$, known as the K-$\Gamma$ model. This material has a honeycomb lattice, and is expected to be a representative of the Kitaev materials (which can realize the 2d Kitaev model). However, the behaviour of RuCl$_3$ shows that the exchange interactions of the material are not purely Kitaev-like, especially because it has an antiferromagnetic
ordering at low temperatures and under low strengths of an externally applied magnetic field. Fitting the data obtained from the measurements of the magnetotropic coefficient (the thermodynamic coefficient associated with magnetic anisotropy),
reported in \textit{Nature Physics 17, 240–244 (2021)}, we estimate the values of the exchange couplings of the effective Hamiltonian. The fits indicate that the Kitaev couplings are subdominant to the other exchange couplings.
\end{abstract}


\maketitle

%==========================================================================

%%%%%%%%%%%%%%%%%%%%%%%%
\section{Introduction}

Spin-orbit coupling (SOC) assisted (spin $j= 1/2$) Mott insulators, exhibiting bond-directional exchange interactions, are expected to exhibit unconventional quantum magnetic phases like spin liquids~\cite{balents-nature,qsl-review-lucile}, predicted by the two-dimensional (2d) Kitaev model~\cite{kitaev} on honeycomb lattice. These putative quantum spin liquids are dubbed as ``Kitaev spin liquids'' (KSLs)~\cite{jackeli,rau,kee,winter,trebst}, and the materials expected to show such behaviour are called Kitave materials. Compounds like honeycomb iridates and $\alpha$-RuCl$_3$ have been identified as candidate Kitaev materials. The hallmark feature of a Kitaev material is that the Kitaev coupling is the dominant exchange coupling.
However, the behaviour of RuCl$_3$ shows that the exchange interactions of the material are not purely Kitaev-like. In this paper, we will address the unresolved question regarding what the possible exchange couplings in $\alpha$-RuCl$_3$ \cite{zigzag3,winter,sizyuk,trebst,kira} could be -- for example, what the dominant terms in the effective spin Hamiltonian are, and whether we can estimate the values of these coupling constants. 

%%%%%%%%%%%%
At low energies, experiments~\cite{nagler2016,nagler2018,muon-zigzag} show signatures consistent with a zig-zag antiferromagnet (AFM) background (also consistent with {\it ab initio} calculations~\cite{zigzag1,zigzag2,zigzag3}), while indicating the existence of an unconventional quantum magnetic phase, which could be the much sought-after KSL induced by a finite magnetic field. 
Exact numerical diagonalization methods to investigate the data from dynamical spin structure factors, and that from heat capacity measurements~\cite{do,leahy}, found that off-diagonal interactions are dominant rather than Kitaev interactions~\cite{suzuki}. On the other hand, other computational papers~\cite{zigzag3,winter,sizyuk,Chaloupka} report that Kitaev terms are the dominant ones.
%%%%%%%%%%%%%%
We focus on the results from resonant torsion magnetometry experiments~\cite{brad-rucl}, which can measure the magnetotropic coefficient $k \equiv \frac{\partial^2 F}{\partial \theta^2}$ at temperature $\mathcal {T}$. Here, $F= -\beta^{-1}\ln Z$ is the free energy, $\beta =\frac{1}{k_B \mathcal {T}}$, and $\theta$ is the angle between the applied magnetic field $\vec B $ and the $c$-axis of the crystal. Using a simple Hamiltonian with a dominant paramagnetic term, we will show that we can fit the data obtained from the measurements of the magnetotropic coefficient, and the fits correspond to the Kitaev terms being subdominant in the so-called K-$\Gamma$ model.

%%%%%%%%%%%%%%%%%%%%%%%%%%%%%%%%%%%%%%%%%%%
\section{Model}

Due to the presence of on-site SOC, the effective magnetic field components along the spin projections are given by:
\begin{align}
{\tilde B}_\alpha & \equiv B_{\gamma}\,D_{\gamma \alpha} \,,\quad
%%%%%%%%%%%%%%%
[D]  = \mathcal{A}\, \mathbb{1}_{3\times 3}+ \begin{pmatrix}
0 &  \mathcal{B} &  \mathcal{B}  \\
%%%%%%%%%%%%%%%%
\mathcal{B} & 0 & \mathcal{B} \\
%%%%%%%%%%%%%%%%%%%%%%%%%%
\mathcal{B} & \mathcal{B} & 0
\end{pmatrix} ,
 \label{D-mat}
\end{align}
where the form of $[D]$ has been fixed by the $C_3$ and $C_2$ rotation symmetries~\cite{crystal-sym} of P$3_112$, constraining it to have $\mathcal{A}$ and $\mathcal{B}$ as the only two independent components (see Appendix \ref{app1}).
In the $abc$-coordinate system, $[D]$ is rotated to take the diagonal form, $\text{diag} \lbrace g_a, g_a , g_c\rbrace$, where the $g$-factors are given by:
\begin{align}
& \tilde g_a = \frac{k_B \, g_a} {\mu_B}\,,\quad 
\tilde g_c = \frac{k_B \,g_c } {\mu_B}\,,
\quad
g_a =   \mathcal{A}- \mathcal{B}\,,\quad
g_c = \mathcal{A} +2\, \mathcal{B}\,,
\end{align}
such that $ k_B = 1.38064852 \times 10^{-23} J/K $ and $ \mu_B=9.274\times 10^{-24}\, J/T$. 
The SOC thus forces the leading order paramegnetic term in our model Hamiltonian to be
$H_0  =  - \sum \limits_{\alpha = \lbrace x, y, z
\rbrace} \tilde B_\alpha \, \sigma_j^\alpha $, rather than $ \left (- \sum \limits_{\alpha = \lbrace x, y, z
\rbrace}  B_\alpha \, \sigma_j^\alpha \right )$.
%%%%%%%%%%%%%%%%%%%%%%
We note that the Hamiltonian has the units of $g\,\mu_B\, \boldsymbol{\sigma} \cdot \vec B$ ($\boldsymbol{\sigma} $ is the dimensionless spin-1/2 vector operator), such that $\frac{g\,\mu_B\, \boldsymbol{\sigma} \cdot \vec B} { k_B \, \mathcal T}$ is dimensionless (because we have factors like $e^{-\beta  H}$). Hence, $\mathcal{A}$ and $\mathcal{B}$ have units of $K/T$.

Following the arguments above, the physics of a honeycomb lattice, restricted to nearest-neighbor interactions, and subjected to an external magnetic field $\vec B $, can be captured by a Hamiltonian of the form: 
\begin{align}
H = H_0 +  V\,,\quad
  H_0  = \, - \sum \limits_{\alpha = \lbrace x, y, z\rbrace} 
  \tilde B_\alpha \,\sigma_j^\alpha \,,\quad
 V =\sum \limits_{\gamma = \lbrace x, y, z\rbrace}
 \sum \limits_{\langle j k\rangle_{\gamma-links} } J^\gamma_{\alpha\,\beta}\,\sigma_j^\alpha  \,\sigma_k^\beta \,,
\end{align}
where $H_0$ is the leading order part for large $B$, and $V$ is the subleading part.
The second summation in $V$ runs over nearest-neighbour spins at sites $j$ and $k$, coupled by a bond along the $\gamma = (x,y,z)$ direction.
%%%%%%%%%%%%%%%
Furthermore, $J^\gamma_{\alpha\,\beta} $ is the coupling constant (for a given value of $\alpha$, $\beta$, and $\gamma$),
$\sigma^{\mu}_j \,(\mu =x,y,z)$ is the Pauli spin matrix representing the spin-$1/2$ operator on site $j$, projected along the $\mu$-axis. The spins are located on the vertices of the honeycomb lattice and $\langle j k\rangle $ denotes the labels of the nearest-neighbour spins. For a hexagonal lattice, there are three different kinds of bonds that can be grouped according to their alignments (see Fig.~3 of Ref.~\cite{kitaev}) -- vertical bonds (which we label as $z$-links), bonds with positive slope (which we label as $x$-links), and bonds with negative slope (which we label as $y$-links). Hence, a given site $j$ is connected to three other sites by these three different types of links, 
denoted by $\gamma$. In other words, the links are not oriented along the 3d orthogonal Cartesian coordinate directions, but each $\gamma$-value denotes the orientation of the bond we are referring to.

The parameters $J^\gamma_{\alpha\,\beta} $ are obtained from the K-$\Gamma$ model, where K stands for the Kitaev term of the 2d Kitaev model~\cite{kitaev}, and $\Gamma$ represents the off-diagonal exchange interactions~\cite{winter,sizyuk,Chaloupka,trebst}, as follows:
%%%%%%%
\begin{align}
H_K &=    \kappa \left ( \sum_{\langle j, k\rangle _{x\textbf{-links}}}
\sigma_j^x\,\sigma_k^x
+ \sum_{ \langle j, k\rangle _{y\textbf{-links}} } \sigma_j^y\,\sigma_k^y
+ \sum_{ \langle j, k\rangle_{z\textbf{-links}}}
\sigma_j^z\,\sigma_k^z \right ),\nn
%%%%%%%%%%%%%%%%%%%%%%%%%%
H_\Gamma & =
\Gamma \left [ \sum_{ \langle j, k\rangle_{x\textbf{-links}}} \left (\sigma_j^y\,\sigma_k^z
+ \sigma_j^z\,\sigma_k^y \right )
+ \sum_{ \langle j, k\rangle_{y\textbf{-links}}} \left (\sigma_j^z\,\sigma_k^x
+ \sigma_j^x \,\sigma_k^z \right )
+  \sum_{\langle j, k\rangle_{z\textbf{-links}}} \left (\sigma_j^x\,\sigma_k^y
+ \sigma_j^y \,\sigma_k^x \right )
  \right ]\nn
  %%%%%%%%%%%%%%%%%%
 &  \quad +
\mathcal{D} \left [ \sum_{ \langle j, k\rangle_{x\textbf{-links}}} \left (\sigma_j^y\,\sigma_k^z
- \sigma_j^z\,\sigma_k^y \right )
+ \sum_{ \langle j, k\rangle_{y\textbf{-links}}} \left (\sigma_j^z\,\sigma_k^x
- \sigma_j^x \,\sigma_k^z \right )
+  \sum_{\langle j, k\rangle_{z\textbf{-links}}} \left (\sigma_j^x\,\sigma_k^y
- \sigma_j^y \,\sigma_k^x \right )
  \right ].
\end{align}
Here $\kappa$ is the strength of the Kitaev term, and $\Gamma$ represents the off-diagonal exchange interaction strengths. The parameters $\Gamma$ and $\mathcal{D}$ represent the symmetric and antisymmetric parts of the off-diagonal terms. 

%%%%%%%%%%%%%%%%%%%%%%%
%%%%%%%%%%%%%%%%%%%%%%%%%%%%%
\begin{figure}[]
\centering
 \includegraphics[width=0.35 \textwidth]{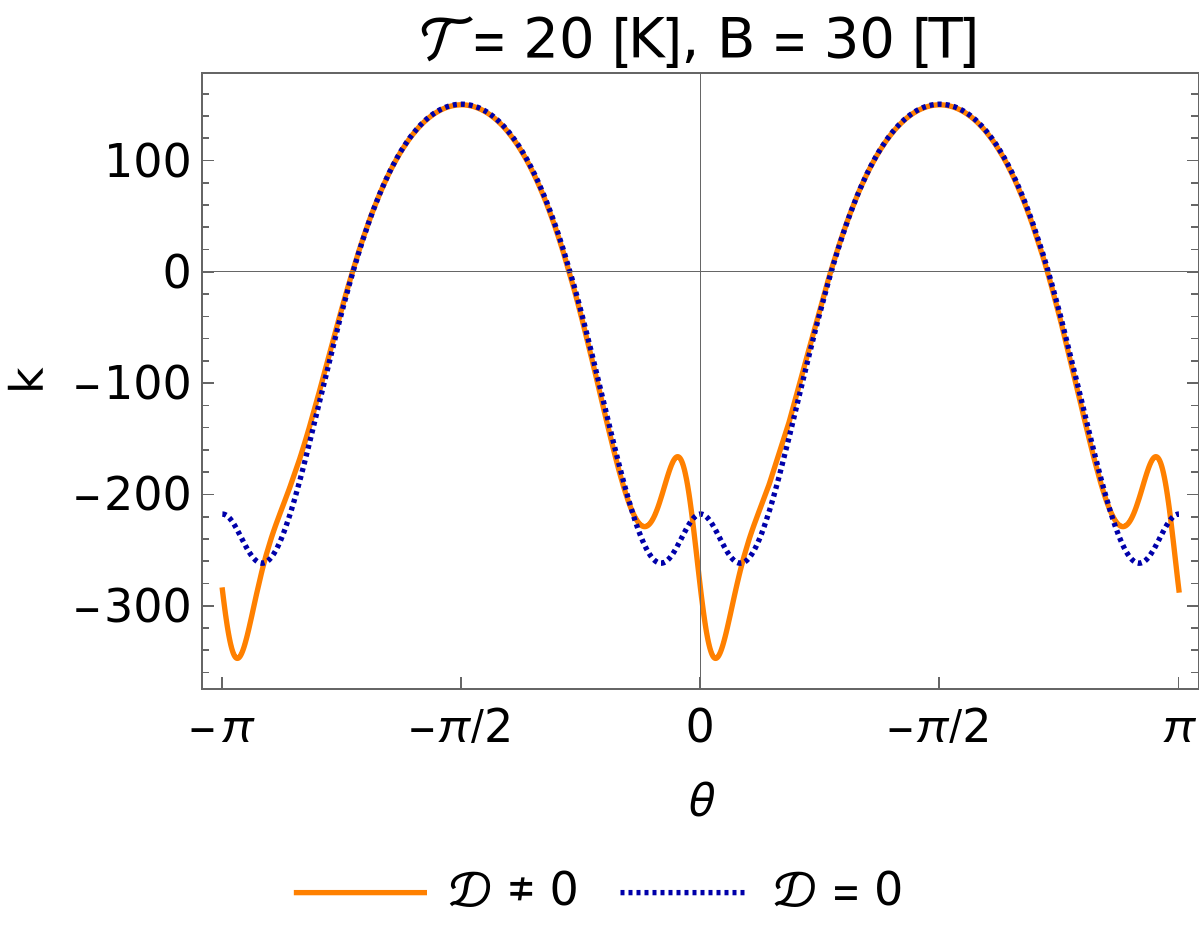}
 \caption{\label{figcompare}
 Plots of $k$ versus $\theta$ computed from our theoretical model, where the orange curve represents the one obtained with the best-fit parameters for the $\left( \mathcal {T}= 20\,K,\, B=30\,T \right ) $ data-set, whereas the dotted blue curve has been drawn using those same parameters, except that we have set $\mathcal{D}=0$. This clearly shows that we can never get an asymmetric spike around $\theta =0$ without an asymmetric off-diagonal $\Gamma$ term, which represents DM interactions.}
\end{figure}
%%%%%%%%%%%%%%%%%%%%%%%%%%

If the trivial paramagnetic term $H_0$ dominates the response of the system to the external magnetic field, we can perform a thermodynamic expansion of the partition function~\cite{book}, as reviewed in Appendix~\ref{app2}. 
Then the partition function, corrected to leading order, evaluates to:
\begin{align}
\mathcal{Z}(T)
&=
 \left [ 2\,\cosh(\beta\,\tilde B )  \right ]^{2N_c} 
\Big[ 1 +
\beta \, N_c \sum \limits_{\gamma, \,\alpha'  , \,\, \lambda'}
  J^\gamma_{\alpha'\,\lambda' }   \,
\frac{ 
\sinh \left(\beta \,\tilde B_{\alpha'} \right)\, \sinh \left(\beta \,\tilde B_{\lambda'}\right)
}
{\cosh^2 \left (\beta \,\tilde B \right)} 
\Big]\,,\quad
\text{ where  }  \tilde B = \sqrt{\sum \limits _\alpha \tilde B _\alpha\, \tilde B_\alpha }
\,,
\label{eqzthermo}
\end{align}
where $N_c$ is the number of unit cells (or, half the number of honeycomb lattice sites) in the system.
It turn outs that this simple model can indeed explain the experimental data, to a high degree of precision.

%%%%%%%%%%%%%%%%%%%%%%%%%%%%%%%%
\section{Fitting the Data}

%%%%%%%%%%%%%%%%%%%%%%%%%%%%%
\begin{figure}[]
\centering
 \includegraphics[width=0.75 \textwidth]{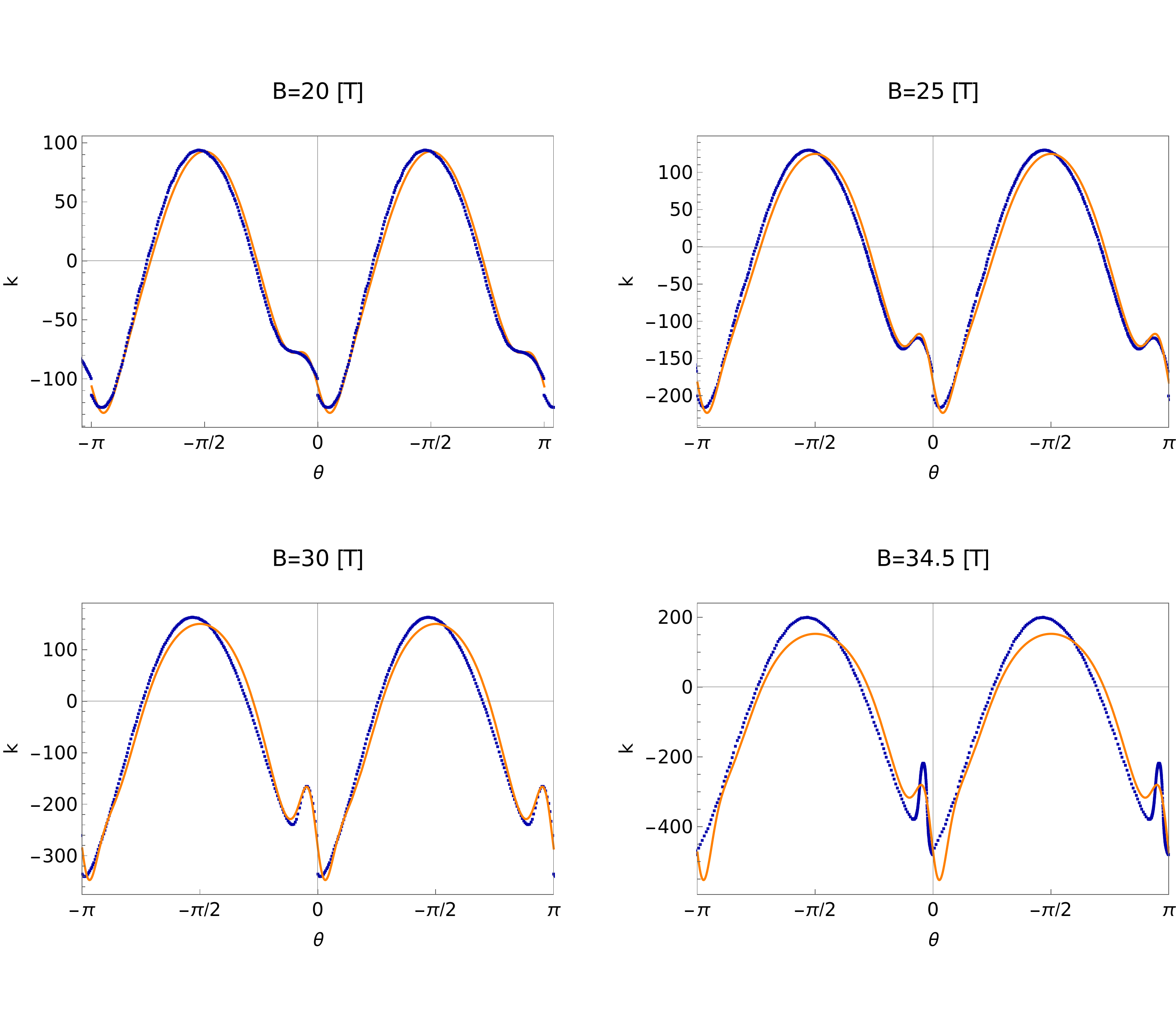}
 \caption{\label{figg1}The data-sets for $k$ versus $\theta$ (in radians) at $\mathcal {T}=20\,K$ for various values of the applied magnetic field strength $B$ (in units of Tesla). We have represented the experimental data-points in blue, and the best-fit curves in orange.}
\end{figure}

%%%%%%%%%%%%%%%%%%%%
\begin{table}[]
\begin{tabular}{ l l l c c c}
\hline
\color{Sepia} $B= 20 \,T $ & & \vline & \textbf{\color{NavyBlue}{Estimate}}
& \textbf{\color{Periwinkle}{Standard Error}} & 
\textbf{\color{MidnightBlue}{Confidence Interval}} \\
\hline
\color{NavyBlue} $\kappa$ & & \vline & 54.4	& 2.37 &	\{52.1, 56.7\}\\
\color{NavyBlue} $\Gamma$ & & \vline & 102 &	6.09	& \{96.8, 108\}\\
\color{NavyBlue} $\mathcal{D}$ & & \vline & -9.60 &	0.629	& \{-10.2, -8.98\} \\
\color{NavyBlue} $g_a$ & & \vline & 4.00 &	0.137	& \{3.87, 4.13\} \\
\color{NavyBlue} $g_c$  &  &\vline & 1.79	& 0.0984 & \{1.70, 1.89\} \\
\color{NavyBlue} $\zeta$ & & \vline & 2.00	& 0.0736	& \{1.93, 2.07\} \\
\color{NavyBlue} $\eta$ & & \vline & -9.38	& 0.150	& \{-9.53, -9.23\} \\
%%%%%%%%%%
\hline
\end{tabular}
\qquad 
%%%%%%%%%%%%%%%%%%%%
\begin{tabular}{ l l l c c c}
\hline
\color{Sepia} $B= 25 \,T $  & & \vline & \textbf{\color{NavyBlue}{Estimate}}
& \textbf{\color{Periwinkle}{Standard Error}} & 
\textbf{\color{MidnightBlue}{Confidence Interval}} \\
\hline
\color{NavyBlue} $\kappa$ & & \vline & 
74.6 & 1.32 & \{73.3, 75.9\}\\
\color{NavyBlue} $\Gamma$ & & \vline & 
133	& 8.80	& \{125,142\}\\
\color{NavyBlue} $\mathcal{D}$ & & \vline & 
-13.9 &	0.866 &	\{-14.7,-13.1\} \\
\color{NavyBlue} $g_a$ & & \vline & 4.00 &	0.124 &	\{3.88, 4.12\} \\
\color{NavyBlue} $g_c$  &  &\vline & 1.52	& 0.0677 &	\{1.46, 1.59\} \\
\color{NavyBlue} $\zeta$ & & \vline & 2.00	& 0.071 &	\{1.93, 2.07\} \\
\color{NavyBlue} $\eta$ & & \vline & -24.4	& 0.281	& \{-24.7, -24.2\} \\
%%%%%%%%%%
\hline
\end{tabular}
%%%%%%%%%%%%%%%%%%%%%%%
\\ \vspace*{0.25 cm}
%%%%%%%%%%%%%%%%%%%%
\begin{tabular}{ l l l c c c}
\hline
\color{Sepia} $B= 30 \,T $  & & \vline & \textbf{\color{NavyBlue}{Estimate}}
& \textbf{\color{Periwinkle}{Standard Error}} & 
\textbf{\color{MidnightBlue}{Confidence Interval}} \\
\hline
\color{NavyBlue} $\kappa$ & & \vline & 
95.8 &	1.46 &	\{94.4, 97.3\} \\
\color{NavyBlue} $\Gamma$ & & \vline & 
152	& 9.46 &	\{143, 161\}\\
\color{NavyBlue} $\mathcal{D}$ & & \vline & 
-15.8 &	0.841	&\{-16.6, -15.0\} \\
\color{NavyBlue} $g_a$ & & \vline &
4.00 &	0.118 &	\{3.88, 4.11\} \\
\color{NavyBlue} $g_c$  &  &\vline & 
1.26 & 0.0486 & \{1.21, 1.30\} \\
\color{NavyBlue} $\zeta$ & & \vline & 2.00 & 0.0687	& \{1.93, 2.07\} \\
\color{NavyBlue} $\eta$ & & \vline & 
-50.1 &	0.507	& \{-50.6, -49.6\} \\
%%%%%%%%%%
\hline
\end{tabular}
%%%%%%%%%%%%%%%%%%%%%%%
\qquad  
%%%%%%%%%%%%%%%%%%%%%%%
%%%%%%%%%%%%%%%%%%%%
\begin{tabular}{ l l l c c c}
\hline
\color{Sepia} $B= 34.5 \,T $  & & \vline & \textbf{\color{NavyBlue}{Estimate}}
& \textbf{\color{Periwinkle}{Standard Error}} & 
\textbf{\color{MidnightBlue}{Confidence Interval}} \\
\hline
\color{NavyBlue} $\kappa$ & & \vline & 
85.5	& 2.90	& \{82.6, 88.3\}
\\
\color{NavyBlue} $\Gamma$ & & \vline & 
156	& 28.7 &	\{128, 184\}
\\
\color{NavyBlue} $\mathcal{D}$ & & \vline & 
-20.3	& 3.31 &	\{-23.5, -17.1\}
 \\
\color{NavyBlue} $g_a$ & & \vline & 4.00 & 0.276	& \{3.73, 4.27\}
 \\
\color{NavyBlue} $g_c$  &  &\vline & 1.20	& 0.122	& \{1.08, 1.32\} \\
\color{NavyBlue} $\zeta$ & & \vline & 2.00 &	0.145	& \{1.86, 2.14\}
 \\
\color{NavyBlue} $\eta$ & & \vline & -88.8	& 1.42	& {-90.2, -87.4}
 \\
%%%%%%%%%%
\hline
\end{tabular}
%%%%%%%%%%%%%%%%%%%%%%%
\caption{\label{figg2}
The table shows the fitting of parameters at $67$\% confidence level, for the $k$ versus $\theta$ data-set. Here, $B$ and $\lbrace   \kappa, \Gamma, \mathcal{D}\rbrace $ are in units of $T$, $\lbrace g_a, g_c\rbrace $ are in units of $K/T$, $\zeta$ is dimensionless, and $\eta$ has the same unit as $k$.}
\end{table}

%%%%%%%%%%%%%%%%%%%%%%%%%%%%%%%
%\begin{figure}[]
%\centering
%\includegraphics[width=0.3 \textwidth]{table}
% \caption{\label{figg2}
%The table shows the fitting of parameters at $67$\% confidence level, for the $k$ versus $\theta$ data-set. Here, $B$ and $\lbrace   \kappa, \Gamma, \mathcal{D}\rbrace $ are in units of $T$, $\lbrace g_a, g_c\rbrace $ are in units of $K/T$, $\zeta$ is unitless, and $\eta$ has the same unit as $k$.}
%\end{figure}
%%%%%%%%%%%%%%%%%%%%%%%%%%%%%%%%%%%%%%%%%%%

%%%%%%%%%%%%%%%%%%%%%%%%%%%%%
\begin{figure}[]
\centering
 \includegraphics[width=0.45 \textwidth]{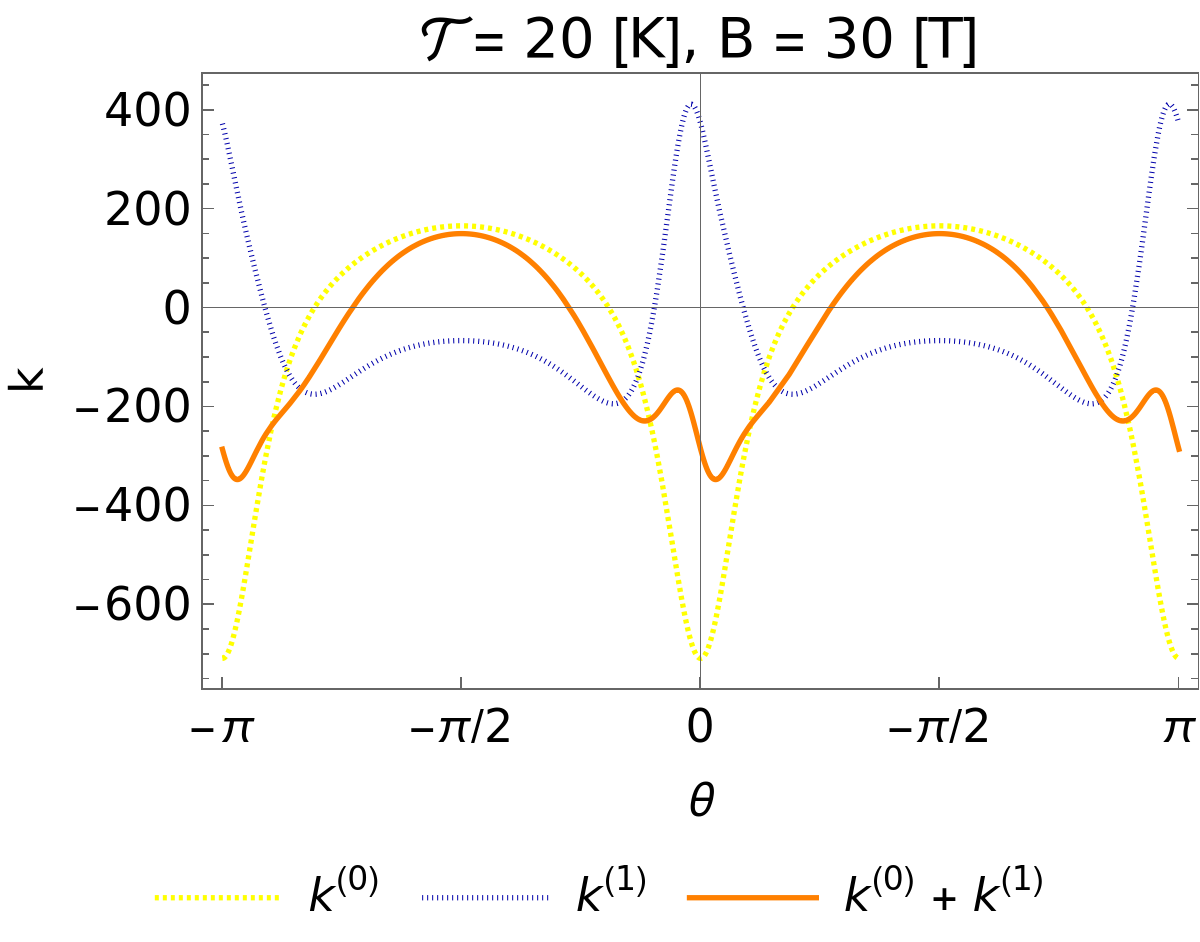}
 \caption{\label{figcor}
We have plotted three curves corresponding to (1) leading order expression $k^{(0)} $; (2) first order correction  $k^{(1)}$; and
(3) $k^{(0)} + k^{(1)}$, as functions of $\theta $, which have been computed from our theoretical model, using the best-fit parameters for the $\left( \mathcal {T}= 20\,K,\, B=30\,T \right ) $ data-set. These three curves are shown in dashed yellow, dotted blue, and orange, respectively.}
\end{figure}
%%%%%%%%%%%%%%%%%%%%%%%

%%%%%%%%%%%%%%%%%%%%%%%%%%%%%
\begin{figure}[]
\centering
 \includegraphics[width=0.45 \textwidth]{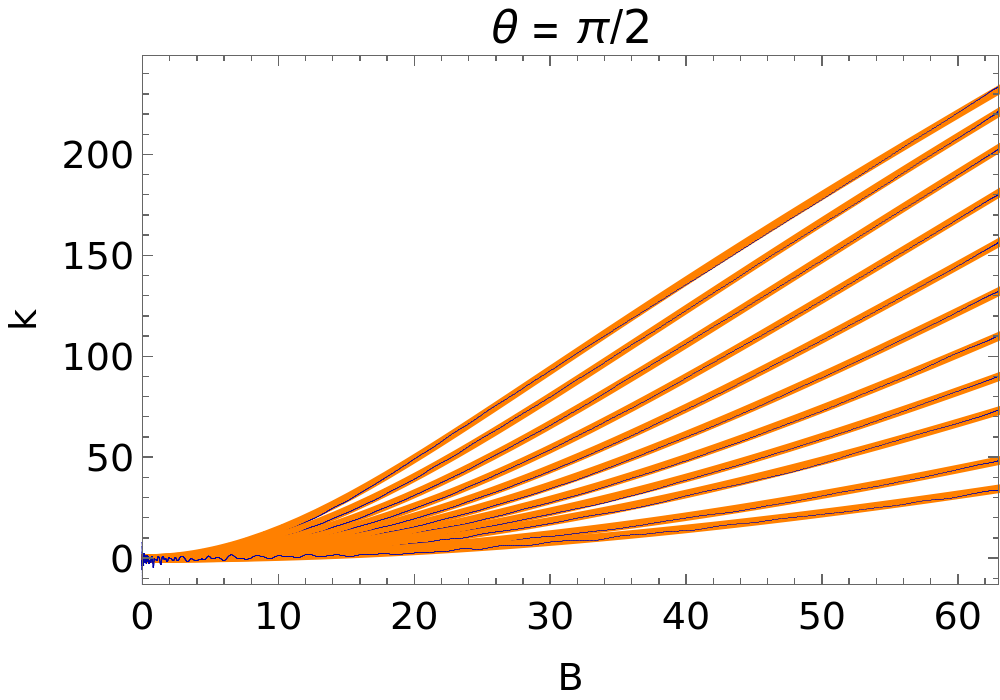}
 \caption{\label{figg3}The data-sets for $k$ versus $B$ (in units of Tesla) at $\theta =\pi/2$ for temperatures ranging from $\mathcal{T} = 30 \,K$ to $\mathcal T= 150 \, K$ at intervals of $10 \,K$. The topmost curve represents the $30\, K$ data, whereas the lowermost curve represents the $150 \, K$ data. We have represented the experimental data-points in blue, and the best-fit curves in orange.}
\end{figure}
%%%%%%%%%%%%%%%%%%%%%%%%%%

%%%%%%%%%%%%%%%%%%%%%%%
\begin{table}
%%%%%%%%%%%%%%%%%%%%
\begin{tabular}{ l l l c c c}
\hline
\color{Sepia} $\mathcal T= 30\,K $  & & \vline & \textbf{\color{NavyBlue}{Estimate}}
& \textbf{\color{Periwinkle}{Standard Error}} & 
\textbf{\color{MidnightBlue}{Confidence Interval}} \\
\hline
\color{NavyBlue} $g_a$ & & \vline & 
2.43 &	0.00172	& \{2.43, 2.44\}
 \\
\color{NavyBlue} $g_c$  &  &\vline & 1.20 &	0.00148	& \{1.2, 1.20\}\\
\color{NavyBlue} $\eta$ & & \vline & 0.000131 & 0.0158	& \{-0.0152, 0.0155\}
 \\
%%%%%%%%%%
\hline
\end{tabular}  \qquad  
%%%%%%%%
%%%%%%%%%%%%%%%%%%%%
\begin{tabular}{ l l l c c c}
\hline
\color{Sepia} $\mathcal T= 40\,K $  & & \vline & \textbf{\color{NavyBlue}{Estimate}}
& \textbf{\color{Periwinkle}{Standard Error}} & 
\textbf{\color{MidnightBlue}{Confidence Interval}} \\
\hline
\color{NavyBlue} $g_a$ & & \vline & 
2.46 &	0.000785	& \{2.46, 2.46\}
 \\
\color{NavyBlue} $g_c$  &  &\vline & 1.24 &	0.000738 & \{1.24, 1.24\}
\\
\color{NavyBlue} $\eta$ & & \vline & 
0.000267 &	0.00700 & 	\{-0.00655, 0.00709\}
 \\
%%%%%%%%%%
\hline
\end{tabular}
%%%%%%%%%%%%%%%%%%%%%%%
%%%%%%%%%%%%%%%%%%%%%%%
\\ \vspace*{0.25 cm}
%%%%%%%%%%%%%%%%%%%%
\begin{tabular}{ l l l c c c}
\hline
\color{Sepia} $\mathcal T= 50\,K $  & & \vline & \textbf{\color{NavyBlue}{Estimate}}
& \textbf{\color{Periwinkle}{Standard Error}} & 
\textbf{\color{MidnightBlue}{Confidence Interval}} \\
\hline
\color{NavyBlue} $g_a$ & & \vline & 
2.42 & 0.000581 & \{2.42, 2.42\}
 \\
\color{NavyBlue} $g_c$  &  &\vline & 
1.23 & 0.000600 & \{1.23, 1.23\},
\\
\color{NavyBlue} $\eta$ & & \vline & 
0.00822 & 0.00473 & \{0.00362, 0.0128\}
 \\
%%%%%%%%%%
\hline
\end{tabular}
 \qquad 
%%%%%%%%
%%%%%%%%%%%%%%%%%%%%
\begin{tabular}{ l l l c c c}
\hline
\color{Sepia} $\mathcal T= 60\,K $  & & \vline & \textbf{\color{NavyBlue}{Estimate}}
& \textbf{\color{Periwinkle}{Standard Error}} & 
\textbf{\color{MidnightBlue}{Confidence Interval}} \\
\hline
\color{NavyBlue} $g_a$ & & \vline & 
2.34 &	0.000456 &	\{2.34, 2.34\}
 \\
\color{NavyBlue} $g_c$  &  &\vline & 1.20 &	0.000518	& \{1.20, 1.20\}
\\
\color{NavyBlue} $\eta$ & & \vline & 
0.00592	&0.00310	&\{0.00290, 0.00894\}\\
%%%%%%%%%%
\hline
\end{tabular}
%%%%%%%%%%%%%%%%%%%%%%%
%%%%%%%%%%%%%%%%%%%%%%%
\\ \vspace*{0.25 cm}
%%%%%%%%%%%%%%%%%%%%
\begin{tabular}{ l l l c c c}
\hline
\color{Sepia} $\mathcal T= 70\,K $  & & \vline & \textbf{\color{NavyBlue}{Estimate}}
& \textbf{\color{Periwinkle}{Standard Error}} & 
\textbf{\color{MidnightBlue}{Confidence Interval}} \\
\hline
\color{NavyBlue} $g_a$ & & \vline & 
 2.28 & 0.000920 & \{2.28, 2.28\} \\
\color{NavyBlue} $g_c$  &  &\vline & 
1.20 & 0.00113 & \{1.20, 1.20\} \\
\color{NavyBlue} $\eta$ & & \vline & 
0.00923 & 0.00490 & \{0.00446, 0.0140\}\\
%%%%%%%%%%
\hline
\end{tabular}
 \qquad 
%%%%%%%%%%%%%%%%%%%%
\begin{tabular}{ l l l c c c}
\hline
\color{Sepia} $\mathcal T= 80\,K $  & & \vline & \textbf{\color{NavyBlue}{Estimate}}
& \textbf{\color{Periwinkle}{Standard Error}} & 
\textbf{\color{MidnightBlue}{Confidence Interval}} \\
\hline
\color{NavyBlue} $g_a$ & & \vline & 
 2.20 & 0.00158 & \{2.2, 2.2\}  \\
\color{NavyBlue} $g_c$  &  &\vline &  1.20 & 0.00204 & \{1.20, 1.20\} \\
\color{NavyBlue} $\eta$ & & \vline & 
 0.0452 & 0.00637 & \{0.039, 0.0514\}\\
%%%%%%%%%%
\hline
\end{tabular}
%%%%%%%%%%%%%%%%%%%%%%%
%%%%%%%%%%%%%%%%%%%%%%%
\\ \vspace*{0.25 cm}
%%%%%%%%%%%%%%%%%%%%
%%%%%%%%%%%%%%%%%%%%
\begin{tabular}{ l l l c c c}
\hline
\color{Sepia} $\mathcal T= 90\,K $  & & \vline & \textbf{\color{NavyBlue}{Estimate}}
& \textbf{\color{Periwinkle}{Standard Error}} & 
\textbf{\color{MidnightBlue}{Confidence Interval}} \\
\hline
\color{NavyBlue} $g_a$ & & \vline & 
 2.12 & 0.00235 & \{2.12, 2.12\} \\
\color{NavyBlue} $g_c$  &  &\vline & 
1.20 & 0.00316 & \{1.20, 1.20\} \\
\color{NavyBlue} $\eta$ & & \vline & 
 -0.0124 & 0.00691 & \{-0.0191, -0.00565\\
%%%%%%%%%%
\hline
\end{tabular}
%%%%%%%%%%
\qquad 
%%%%%%%%%%%%%%%%%%%%
\begin{tabular}{ l l l c c c}
\hline
\color{Sepia} $\mathcal T= 100\,K $  & & \vline & \textbf{\color{NavyBlue}{Estimate}}
& \textbf{\color{Periwinkle}{Standard Error}} & 
\textbf{\color{MidnightBlue}{Confidence Interval}} \\
\hline
\color{NavyBlue} $g_a$ & & \vline & 
 2.03 & 0.00351 & \{2.03, 2.03\} \\ 
\color{NavyBlue} $g_c$  &  &\vline &  
1.20 & 0.00481 & \{1.20, 1.20\} \\
\color{NavyBlue} $\eta$ & & \vline & 
0.043 & 0.00740 & \{0.0358, 0.0503\} \\
%%%%%%%%%%
\hline
\end{tabular}
%%%%%%%%%%%%%%%%%%%%%%%
%%%%%%%%%%%%%%%%%%%%%%%
\\ \vspace*{0.25 cm}
%%%%%%%%%%%%%%%%%%%%
%%%%%%%%%%%%%%%%%%%%
\begin{tabular}{ l l l c c c}
\hline
\color{Sepia} $\mathcal T= 110\,K $  & & \vline & \textbf{\color{NavyBlue}{Estimate}}
& \textbf{\color{Periwinkle}{Standard Error}} & 
\textbf{\color{MidnightBlue}{Confidence Interval}} \\
\hline
\color{NavyBlue} $g_a$ & & \vline & 
 1.94 & 0.00531 & \{1.94, 1.95\} \\
\color{NavyBlue} $g_c$  &  &\vline & 
 1.20 & 0.00732 & \{1.19, 1.201\} \\
\color{NavyBlue} $\eta$ & & \vline & 
 -0.00637 & 0.0078 & \{-0.014, 0.00123\}\\
%%%%%%%%%%
\hline
\end{tabular}
%%%%%%%%%%
\qquad 
%%%%%%%%%%%%%%%%%%%%
\begin{tabular}{ l l l c c c}
\hline
\color{Sepia} $\mathcal T= 130\,K $  & & \vline & \textbf{\color{NavyBlue}{Estimate}}
& \textbf{\color{Periwinkle}{Standard Error}} & 
\textbf{\color{MidnightBlue}{Confidence Interval}} \\
\hline
\color{NavyBlue} $g_a$ & & \vline & 
1.79 & 0.0111 & \{1.78, 1.80\} \\
\color{NavyBlue} $g_c$  &  &\vline &  
1.20 & 0.0151 & \{1.18,1.21\} \\
\color{NavyBlue} $\eta$ & & \vline & 
 -0.0081 & 0.00815 & \{-0.016,-0.000166\}  \\
%%%%%%%%%%
\hline
\end{tabular}
%%%%%%%%%%%%%%%%%%%%%%%
%%%%%%%%%%%%%%%%%%%%%%%
\\ \vspace*{0.25 cm}
%%%%%%%%%%%%%%%%%%%%
%%%%%%%%%%%%%%%%%%%%
\begin{tabular}{ l l l c c c}
\hline
\color{Sepia} $\mathcal T= 150\,K $  & & \vline & \textbf{\color{NavyBlue}{Estimate}}
& \textbf{\color{Periwinkle}{Standard Error}} & 
\textbf{\color{MidnightBlue}{Confidence Interval}} \\
\hline
\color{NavyBlue} $g_a$ & & \vline & 
 1.76 & 0.0953 & \{1.66, 1.85\} \\
\color{NavyBlue} $g_c$  &  &\vline & 
1.287 & 0.122 & \{1.17, 1.41\} \\
\color{NavyBlue} $\eta$ & & \vline & 
 -0.00276 & 0.0385 & \{-0.0403, 0.0347\} \\
%%%%%%%%%%
\hline
\end{tabular}
%%%%%%%%%%
\caption{\label{figg4}
The table shows the fitting of parameters at $67$\% confidence level, for the $k$ versus $B$ data-set. Here, temperature $\mathcal {T}$ is in units of Kelvin ($K$), $\lbrace g_a, g_c\rbrace $ are in units of $K/T$, and $\eta$ has the same unit as $k$.}
\end{table}

%%%%%%%%%%%%%%%%%%%%%%%%%%%%%
%\begin{figure}[]
%\centering
% \includegraphics[width=0.3 \textwidth]{table2}
% \caption{\label{figg4}
%The table shows the fitting of parameters at $67$\% confidence level, for the $k$ versus $B$ data-set. Here, temperature $\mathcal {T}$ is in units of Kelvin ($K$), $\lbrace g_a, g_c\rbrace $ are in units of $K/T$, and $\eta$ has the same unit as $k$.}
%\end{figure}
%%%%%%%%%%%%%%%%%%%%%%%%%%%%%%%%%%%%%%%%%%%%

According to some papers in the literature~\cite{zigzag3,cao}, the point-group symmetry of the Ru-Ru links is $C_{2h}$ in a $C/2m$ unit cell, and hence the antisymmetric Dzyaloshinskii-Moriya (DM) exchange is zero. Because spin is an axial vector itself, the non-zero antisymmetric part of $J^\gamma_{\alpha\,\beta} $ is equivalent to $\boldsymbol{P}\cdot \left (\vec S \times \vec S \right )$ term where $\boldsymbol{P}$ is a polar vector. So in order to have a DM term in exchange, the  chemical environment of the Ru-Ru bond must allow for a polar vector. In the undistorted honeycomb lattice, such a polar vector is prohibited by symmetry. It is non-zero for next-nearest neighbor exchange links (even in the undistorted case)~\cite{haldane} or if Cl octahedra are distorted.
%%%%%%%%%%%%%%%%%%%5
 However, the peak shapes in the $k$ versus $\theta $ data (and the behavior in a broader angular range around it) are not symmetric around the $c$-axis, {\it{i.e.}}, $k(\theta) - k(-\theta) \neq 0$. We find that this behavior is possible only if we allow the antisymmetric DM term, which points to a distorted octahedra, leading to a deviation from the assumed crystal symmetries. This is shown in Fig.~\ref{figcompare}, where we show the plots for $k(\theta)$ computed from our model, with the parameter values taken from the best-fit parameters of the $B=30\, T$ data (except that we set $\mathcal{D} = 0$ for the orange curve). 

The expressions for $F$ and $k$ depend on the polar angle $\theta$, but not on the azimuthal angle $\phi$.
 We fit the data-sets for four different values of the applied magnetic field strength $B$, using ``NonlinearModelFit'' of Mathematica. The data-sets for $B\leq 15 \, T$ are not considered as they are either close to or within the AFM phase. Since the scaling and absolute shift of each data-set are uncertain, we include two more parameters, namely, ``$\zeta$'' and ``$\eta$'' corresponding to the unknown scale and shift. The experimental data and the fitted functions are shown in Fig.~\ref{figg1}. The confidence intervals for all the parameters at $67 $\% confidence level are shown in Table~\ref{figg2}.

 We also fit the $k$ versus $ B $ data available for $\theta= \pi/2$. One can check that the correction terms from $J^\gamma_{\alpha\,\beta} $ hardly affect the regions around $\theta= \pi/2$ (see Fig.~\ref{figcor}). They have the most visible impact only around the $\theta= 0$ and $\theta= \pi$ regions. Hence, the fitting process keeping the first order correction makes the parameters indeterminate. However, if we fit only with the zeroth order expression, we get excellent values for $g_a$ and $g_c$. We also need to include a parameter ``$\eta$'' to account for the uncertainty in the absolute shift of the data-set for each temperature value. These fits are shown in Fig.~\ref{figg3}. The data-sets for temperatures $\mathcal{T} \leq 20 \, K$ are not considered as each of them has a considerable region within the AFM phase in the low $B$ ranges, which cannot be fitted by the functional forms meant for the paramagnetic phase. The confidence intervals for $g_a$, $g_c$, and $\eta$, at $67 $\% confidence level, are shown in Table~\ref{figg4}.
 
 %%%%%%%%%%%%%%
\section{Summary and Outlook}
 
Let us discuss some other possibilities which might be responsible for causing the asymmetry in the spike around $\theta=0$ in the $k$ versus $\theta$ data. Firstly, in the experimental set-ups, the path along which the sample is rotated in the external magnetic field to change $\theta$, may deviate from a great circle, leading to an uncertainty of upto $10^{\circ}$. However, incorporating these deviations, the theoretical curves do not show the desired asymmetry~\cite{arkady}.
Secondly, the K-$\Gamma$ Hamiltonian (even without DM, distortion, misalignment of rotation etc.) lacks mirror reflection symmetry. Therefore, magnetotropic coefficients (or free energy) are different for applied magnetic fields, $\vec B$ and $\vec B'$, that are related by a mirror reflection in honeycomb plane. Such asymmetry is artificially removed in low-order perturbation theory. This is directly analogous to accidental symmetries of the standard model (such as separate conservation of baryon ad lepton number) that only exist in the lowest order of expansion in inverse GUT scale. It is a general phenomenon -- low orders in perturbation theory tend to accidentally ``restore'' some of the symmetries of the Hamiltonian. In the Kitaev model, the lowest order perturbation expansion in magnetic field~\cite{kitaev} is symmetric (with respect to mirror-$ab$-plane) -- one needs to go to higher orders in B to see the asymmetry of the Hamiltonian. The same might be true for the thermodynamic perturbative expansion. By going to higher orders (second, or maybe third order), the asymmetric character of the Hamiltonian may eventually show up. However, such higher order computations are beyond the scope of this paper.

Our best-fit parameters show that the Kitaev terms are subdominant to the $\Gamma$ (and $\mathcal{D}$) terms.
In fact, the large $ \Gamma$ value contrasts with the expectation so far~\cite{zigzag3,winter,sizyuk,Chaloupka} that $\alpha$-RuCl$_3$ is a “Kitaev model material”. It has also been predicted~\cite{Chaloupka,zigzag3,majumder} in those models (including a small Heisenberg term) that the ratio $g_c/g_a \simeq 0.4-0.5 $. Our results (see Fig.~\ref{figg2}) are close to these results, although we should remember that our model differs from theirs.

%%%%%%%%%%%%%%%%%%%%%%%%
\section{Acknowledgments}
We thank Michael J. Lawler for suggesting the problem. We are also grateful to Brad Ramshaw, Arkady Shekhter, and Kimberly Modic for insightful discussions.
%------------------------------------------------------------------------------------------------------------------------------------------------------------------------------------------------------------------------

%\bibliography{biblio}

%============================================================================
%.............................................................................
\appendix

%%%%%%%%%%%%%%%%%%%%%%%%%%%%

\section{Choice of coordinate system and crystal symmetries}
\label{app1}

We choose a coordinate system such that the plane of the honeycomb lattice is described by three in-plane vectors $\vec r_1=(0,1,-1), \,\vec  r_2=(-1,1,0), \,\vec r_3 =(-1, 0, 1)$, as they lie on 
the plane formed by cutting the three points: $ (1,0,0), \,(0,1,0) \text{ and }  (0,0,1)$. Then
the perpendicular vector is $\mathbf{r}_{\perp} =(1,1,1)/ \sqrt{3}$ and we choose the in-plane direction as $\vec r_2/\sqrt{2}$, giving
$ \frac{\mathbf{B} } {B} = \frac{ \cos \theta\, (1,1,1)} { \sqrt{3}} + \frac{\sin \theta\, (-1,1,0) } { \sqrt{2}}$,
with the magnetic field making an angle $\theta$ with the $c$-axis.
Let us also define the $a$-axis along the line joining $(1,1,1)/ 3$ and $(1,0,0)$, such that the projection of $\mathbf{B} $ on the $ab$-plane makes an angle $\phi$ with the $a$-axis.
%We choose this coordinate system because the exchange interaction terms take a simple form in this coordinate system, which is captured by Eq.~(6) of the main text. However, the matrix $[D]$ (see Eq.~(2) of main text) is not diagonal for this choice.

Given a unit vector $\mathbf u$, the matrix for a rotation by an angle of $\phi$ about an axis in the direction of $\mathbf u$ is:
 \begin{align}
R(\mathbf u, \phi) &=
\begin{pmatrix}
\cos \phi+ u_x^2  \left( 1-\cos \phi \right) 
& u_x \, u_y   \left( 1-\cos \phi \right) -u_z \sin \phi
& u_x \, u_z   \left( 1-\cos \phi \right) + u_y \sin \phi  \\
%%%%%%%%%%%%%%%%
 u_y \, u_x   \left( 1-\cos \phi \right) + u_z \sin \phi
& \cos \phi+ u_y^2  \left( 1-\cos \phi \right)
& u_y \, u_z   \left( 1-\cos \phi \right) -u_x \sin \phi\\
%%%%%%%%%%%%%%%%%%%%%%%%%%
 u_z \, u_x   \left( 1-\cos \phi \right) -u_y \sin \phi
 & u_z \, u_y   \left( 1-\cos \phi \right) + u_x \sin \phi
& \cos \phi+ u_x^2  \left( 1-\cos \phi \right)  
\end{pmatrix}  .
 \end{align}
 Now the crystal symmetry allows invariance under a $C_3$ rotation about $\mathbf r_{\perp}$, which corresponds to invariance under the rotation matrix:
 \begin{align}
 R(\mathbf r_{\perp}, 2\,\pi/3) 
=   \begin{pmatrix}
0 & 0 & 1   \\
%%%%%%%%%%%%%%%%
1 & 0 & 0 \\
%%%%%%%%%%%%%%%%%%%%%%%%%%
0 & 1 & 0
\end{pmatrix} .
 \label{rot-c3}
\end{align}  

For $C_2$ rotation about $\vec r_1$, we have:
\begin{align}
R \left (\frac{\mathbf r_1} {\sqrt 2}, \pi \right ) &=
\begin{pmatrix}
-1 & 0 &  0  \\
%%%%%%%%%%%%%%%%
  0 & 0 &  -1   \\
%%%%%%%%%%%%%%%%%%%%%%%%%%
0 & -1 & 0
\end{pmatrix}  .
\label{rot-c2}
 \end{align}

Due to on-site spin-orbit coupling, the leading order paramegnetic term in our model is given by
$H_0  =  - \sum \limits_{\alpha = \lbrace x, y, z
\rbrace} \tilde B_\alpha \, \sigma_j^\alpha $, rather than $ \left (- \sum \limits_{\alpha = \lbrace x, y, z
\rbrace}  B_\alpha \, \sigma_j^\alpha \right )$, where $  {\tilde B}_\alpha \equiv B_{\gamma}\,D_{\gamma \alpha}$.
We still have the $C_3$ and $C_2$ rotation symmetries of P$3_112$~\cite{crystal-sym} to be satisfied, which implies that:
\begin{align}
[B]^T\,[D]\,[\sigma] =\left( R\,[B]\,\right)^T \,[D] \, R\,[\sigma]
\Rightarrow D=R^T\,[D] \, R \,,
\end{align}
where $R$ has been defined in Eq.~(\ref{rot-c3}). Then, $R \left (\mathbf r_\perp, 2\,\pi/3 \right )$ and $ R \left (\frac{\mathbf r_1} {\sqrt 2}, \pi \right )$ restrict $[D]$ to have only two independent components, namely $\mathcal{A}$ and $\mathcal{B}$, such that
\begin{align}
[D] = \mathcal{A}\, \mathbb{1}_{3\times 3}+ \begin{pmatrix}
0 &  \mathcal{B} &  \mathcal{B}  \\
%%%%%%%%%%%%%%%%
\mathcal{B} & 0 & \mathcal{B} \\
%%%%%%%%%%%%%%%%%%%%%%%%%%
\mathcal{B} & \mathcal{B} & 0
\end{pmatrix} .
 \label{D-mat}
\end{align}  
 
%%%%%%%%%%%%%%%%%%%%%%%%%%%%%%%
\section{Thermodynamic expansion of the K-$\Gamma$ model in the large magnetic field limit}
\label{app2}

We perform a thermodynamic expansion of the K-$\Gamma$ model in the large magnetic field limit, following the methods describe in Ref.~\onlinecite{book}, which are applicable when we are interested in the thermodynamic properties at finite temperature. We review this perturbation expansion when the
Hamiltonian can be written as $H = H_0 + \lambda \, V$, where $H_0$ is the leading order part for large $B$, and $\lambda $ is the perturbative expansion parameter, with $V$ being the subleading part. 

We are interested in the thermodynamic properties at finite temperature. Thus we start with the canonical partition function:
\begin{align}
\mathcal{Z}(T) =\Tr \left [ e^{-\beta\, H}\right ]
=\Tr \left [ e^{-\beta\left  (  H_0 + \lambda \, V \right ) }\right ],
\end{align}
and seek to expand its logarithm in powers of $\lambda$. Since $H_0$ and $V$ do not commute for the K-$\Gamma$ model, we use the approach employed for interaction picture time evolution. We define the function $f(\beta)$ by:
\begin{align}
& e^{-\beta\left  (  H_0 + \lambda \, V \right ) }=
e^{-\beta\, H_0} \, f(\beta)
%%%%%%%%%%%
\Rightarrow  \frac{df(\beta)}{d\beta}
= -   \lambda \,e^{\beta\, H_0}\, V   e^{-\beta\, H_0} \, f(\beta)\,.
\end{align}
%%%%%%%%%%%%%%%%%%%%%%%%
Casting this in the form of the integral equation, we get:
\begin{align}
f(\beta) &= 1-\lambda \int_0^\beta \,d\tau \,\tilde V(\tau)
 \, f(\tau)\,,\quad
 \tilde V(\tau) = e^{\tau\, H_0}\, V\,   e^{-\tau\, H_0} \,,
\end{align}
which we solve by iteration:
\begin{align}
f(\beta) &=1 +\sum \limits_{n=1}^{\infty}  (-\lambda)^n \int_0^\beta \,d\tau_1
\int_0^{\tau_1 } \,d\tau_2 \cdots\int_0^{\tau_{n-1}} \,d\tau_n\,
\tilde V(\tau_1)  \, \tilde V(\tau_2) \cdots \tilde V(\tau_n)\,.
\end{align}
This gives us the partition function as:
\begin{align}
\mathcal{Z}  (T) =\mathcal{Z}_0
\Big[ 1 + \sum \limits_{n=1}^{\infty}  (-\lambda)^n \int_0^\beta \,d\tau_1
\int_0^{\tau_1 } \,d\tau_2 \cdots\int_0^{\tau_{n-1}} \,d\tau_n\,\langle 
\tilde V(\tau_1)  \, \tilde V(\tau_2) \cdots \tilde V(\tau_n)  \rangle_0 
\Big ]\,,
\end{align}
where $ \langle \cdots \rangle _0$ denotes the unperturbed expectation value:
\begin{align}
 \langle  A \rangle _0 \equiv \frac{ \Tr \left [ e^{ -\beta\, H_0}\, A \right ] }
 {\Tr \left [ e^{ -\beta\, H_0} \right ]}
\end{align}
for any operator $A$. The leading order term is given by:
\begin{align}
 \langle \tilde  V (\tau )\rangle_0 &  =
\frac{ \Tr \left [ e^{ -\beta\, H_0}\, e^{\tau\, H_0}\, V \,  e^{-\tau\, H_0} \right ] }
 {\Tr \left [ e^{ -\beta\, H_0} \right ]}
 = \frac{ \Tr \left [ e^{ -\beta\, H_0}\,   V  \right ] }
 {\Tr \left [ e^{ -\beta\, H_0} \right ]}\,,
 \end{align}
which is in fact independent of $\tau$.

Let us compute the leading term in the partition function for the Hamiltonian of the main text, such that:
\begin{align}
  H_0  = \, - \sum \limits_{\alpha = \lbrace x, y, z
\rbrace} \tilde B_\alpha \, \sigma_j^\alpha \,,\quad
 V =\sum \limits_\gamma \sum \limits_{\langle j k\rangle_{\gamma-links} } J^\gamma_{\alpha\,\beta}\,\sigma_j^\alpha  \,\sigma_k^\beta \,.
\end{align}
%%%%%%%%%%%%%%%%%%%%%%%%%%%%%%%%%%%%%%%%%
Hence, we get:
\begin{align}
 \langle \tilde  V (\tau )\rangle_0
&=\frac{\sum \limits_{\gamma, \,\alpha'  , \,\, \lambda'}\sum \limits_{\langle j l \rangle_{\gamma-links} }
  J^\gamma_{\alpha'\,\lambda' }  
\sinh \left(\beta \, \tilde B_{\alpha'} \right)\, \sinh \left(\beta \,\tilde B_{\lambda'} \right)
}
{\cosh^2 \left (\beta \,\tilde B \right)}
= \frac{N_c  \sum \limits_{\gamma, \,\alpha'  , \,\, \lambda'}
  J^\gamma_{\alpha'\,\lambda' }  
\sinh \left(\beta \,\tilde B_{\alpha'}\right)\, \sinh \left(\beta \,\tilde B_{\lambda'}\right)
}
{\cosh^2 \left (\beta \,\tilde B \right)}
\,,
\end{align}
where $ \tilde B = \sqrt{\sum \limits _\alpha \tilde B _\alpha\, \tilde B_\alpha }\,, $ and $N_c$ is the number of unit cells in the system.
Finally, this gives us the partition function, corrected to leading order, as:
\begin{align}
\mathcal{Z}(T)
&=
 \left [ 2\,\cosh(\beta\,\tilde B)  \right ]^{2N_c} 
\Big[ 1 +
\beta \, N_c \sum \limits_{\gamma, \,\alpha'  , \,\, \lambda'}
  J^\gamma_{\alpha'\,\lambda' }   \,
\frac{ 
\sinh \left(\beta \,\tilde B_{\alpha'} \right)\, \sinh \left(\beta \,\tilde B_{\lambda'}\right)
}
{\cosh^2 \left (\beta \,\tilde B \right)} 
\Big]\,.
\label{eqzthermo}
\end{align}

\bibliography{biblio} 

\end{document}